\documentstyle[12pt]{article}
\newcommand{\sptwo}{1.4}

\newcommand{\doublespace}{\edef\baselinestretch{\sptwo}\Large\normalsize}

\begin{document}
\doublespace
\begin{center}
{\bf Interacting Electrons in Quantum Dots}\\
\renewcommand\thefootnote{\fnsymbol{footnote}}
{Yeong E. Kim \footnote{ e-mail:yekim$@$physics.purdue.edu} and
Alexander L. Zubarev\footnote{ e-mail: zubareva$@$physics.purdue.edu}}\\
Purdue Nuclear and Many-Body Theory Group (PNMBTG)\\
Department of Physics, Purdue University\\
West Lafayette, Indiana  47907\\
\end{center}
\begin{quote}
 The ground states of $N$-electron parabolic quantum dots in 
 the presence of a perpendicular
  magnetic field  are investigated.  Rigorous lower bounds to the ground-state 
energies are obtained.
It is shown that our lower bounds agree well with the results of exact 
diagonalization.
Analytic results for the lower bounds to the ground-state energies of the 
quantum dots in a strong magnetic field
(known as electron molecule) agree very well with numerically calculated lower bounds.
\end{quote} 
\vspace{5mm}
\noindent
PACS: 03.65.Ge, 05.30.Fk, 73.21.La

\vspace{55 mm}

\pagebreak

{\bf I. Introduction}
\vspace{8pt}

In recent years, there has been intense study of nanostructures such as 
quantum dots (QD) [1-5], where  quasi-two-dimensional islands of 
 electrons are
laterally confined by an externally imposed potential that, in a good 
approximation, is 
parabolic. In Ref.[6], the electronic states of interacting 
electrons in three-electrons QD are calculated without making assumptions 
about the shape of the confining potential and dimensionality of the problem.

Theoretical investigations of the ground states of QD have been  reported in many
papers. 
As for the standard Hartree and
Hartree-Fock (HF) approximations, there are doubts about their accuracy, since
the exchange and correlation energies can be significant in QD [7,8].

A simple way to incorporate the interaction between electrons is to use 
the Post model [9], where inter-electron repulsion
 is replaced by the harmonic interaction [10]. For a critical analysis of this
 approximation, see Ref. [11]. The Post model [9] was used 
for a problem in nuclear physics in Ref. [12].

Numerical calculations using exact-diagonalization techniques were
carried out in Refs.[13-19]. These calculations are computationally extensive
and limited to a few ($\le 6$) electrons.

The ground states of an N-electron QD in magnetic fields have been measured up 
to $N\le50$ [20].

The purpose of this work is to provide a rigorous lower bounds to the 
ground-state energy of N-electron QD in magnetic fields for any $N$. We show that our 
lower bounds for ground states agree well with the exact results of the diagonalization method.

The paper is organized as follows. In Section II, we generalize a lower-bound 
method developed by Hall and Post [21] for the case of N-electron QD in a 
magnetic field $B$. In Section III, lower bounds are found analytically in 
the large $B$ limit. In  Section IV, we describe our calculations. A summary 
and conclusions are given in Section V.

\vspace{18pt}
{\bf II. Lower Bounds }
\vspace{8pt}

The Hamiltonian for $N$ interacting electrons confined in a parabolic QD, in 
the presence of a magnetic field $B$ perpendicular to the dot, can be written as
$$
H=\frac{1}{2m^{\ast}}\sum_{i=1}^{N} \vec{p}_i^2+\frac{1}{2}m^{\ast}\Omega^2
\sum_{i=1}^{N} \vec{r}_i^2-\frac{\omega_c}{2}L_z+\sum_{i<j}\frac{e^2}{\epsilon
\mid\vec{r}_i-\vec{r}_j\mid}+g^{\ast}\mu_bBS_z,
\eqno{(1)}
$$
where $m^{\ast}$ is the electron effective mass, 
$\Omega^2=\omega_0^2+\omega_c^2/4$, $\omega_0$ is the parabolic confinement frequency, $\omega_c$ is the cyclotron frequency, 
$L_z$ is the $z$ component of the total orbital momentum, 
$\epsilon$ is the dielectric constant, $g^{\ast}$ is the effective g-factor,
$\mu_b$ is the Bohr magneton, and $S_z$ is the z component of the total spin.

In our numerical calculations we use the effective mass $m^{\ast}=0.067m_e$
($m_e$ is the free-electron mass)
of $GaAs$ QD.

Now we introduce the center-of-mass coordinates, 
$\vec{R}=\frac{1}{N}\sum_{i=1}^{N} \vec{r}_i$ and$~$ 
$\vec{P}=\sum_{i=1}^{N} \vec{p}_i$.

Using 
$$
\sum_{i=1}^{N} \vec{r}_i^2=N\vec{R}^2+\frac{1}{N}\sum_{i<j} (\vec{r}_i-\vec{r}_j)^2,
\eqno{(2)}
$$
$$
\sum_{i=1}^{N} \vec{p}_i^2=\frac{\vec{P}^2}{N}+\frac{1}{N}\sum_{i<j} (\vec{p}_i-\vec{p}_j)^2,
\eqno{(3)}
$$
and
$$
\vec{L}=\sum_{i=1}^{N} \vec{r}_i\times\vec{p}_i=\vec{R}\times\vec{P}+
\frac{1}{N}\sum_{i<j}(\vec{r}_i\times\vec{p}_i+\vec{r}_j\times\vec{p}_j
-\vec{r}_i\times\vec{p}_j-\vec{r}_j\times\vec{p}_i),
\eqno{(4)}
$$
we can rewrite Eq.(1) as
$$
H=H_{cm}+H_{rel}+H_z,
\eqno{(5)}
$$
where the first term is the center-of-mass energy, the second term is 
the relative energy and the last term is the Zeeman energy, with 
$H_z=g^{\ast}\mu_bBS_z$.
$H_{cm}$ and $H_{rel}$ are given by
$$
H_{cm}=\frac{\vec{P}^2}{2m^{\ast}N}+\frac{m^{\ast}N \Omega^2\vec{R}^2}{2}-
\frac{\omega_c}{2}(\vec{R}\times\vec{P})_z,
\eqno{(6)}
$$
and
$$
H_{rel}=\sum_{i<j}H_{ij},
\eqno{(7)}
$$
where
$$
H_{ij}=
\frac{(\vec{p}_i-
\vec{p}_j)^2}{2m^{\ast}N}+\frac{m^{\ast}\Omega^2(\vec{r}_i-\vec{r}_j
)^2}{2N}+\frac{e^2}{\epsilon
\mid\vec{r}_i-\vec{r}_j\mid}-\frac{\omega_c}{2N}(\vec{r}_i\times\vec{p}_i+\vec{r}_j\times\vec{p}_j
-\vec{r}_i\times\vec{p}_j-\vec{r}_j\times\vec{p}_i)_z.
\eqno{(8)}
$$
Hence we have for the ground state energy
$$
E=\hbar\Omega+E_{rel}+g^{\ast}\mu_bBS_z,
\eqno{(9)}
$$
where
$$
E_{rel}=<\psi\mid H_{rel}\mid\psi>,
\eqno{(10)}
$$
and $\psi(\vec{r}_1, \vec{r}_2, ... \vec{r}_N)$ is the ground state wave function
. Using the symmetric properties of $\psi$ we can rewrite Eq. (10) as
$$
E_{rel}=\frac{N(N-1)}{2}<\psi\mid H_{12}\mid\psi>.
\eqno{(11)}
$$
Introducing the Jacobi coordinates $\vec{\zeta}_i$ as
$$
\vec{\zeta}_i=\sum_{j=1}^N U_{ij}\vec{r}_j,
\eqno{(12)}
$$
with
$$
U_{ij}=\left\{\begin{array}{ll}
       (i)^{-1} &\mbox{if $j<i+1,$}\\
       -1          &\mbox{if $j=i+1,$}\\
        0          &\mbox{if $j>i+1,$}
                   \end{array}
\right.
\eqno{(13)}
$$
we have
$$
H_{12}=-\frac{2}{m^{\ast}N}\Delta_{\zeta_1}+ \frac{m^{\ast}\Omega^2\zeta_1^2}
{2N}+\frac{e^2}{\epsilon \zeta_1}-\frac{\omega_c}{N}(\vec{\zeta}_1\times
\frac{\hbar}{i}\nabla_{\zeta_1})_z,
\eqno{(14)}
$$
where $\vec{\zeta}_1=\vec{r}_1-\vec{r}_2$.

Projecting $\mid\psi>$ on the complete basis $\mid n>$, 
generated by the effective two-body Hamiltonian $H_{12}$, 
$~H_{12}\mid n>=E_n \mid n>$, and using
$$<\psi \mid H_{12}\mid \psi>=\sum_n E_n\mid<\psi \mid n><n\mid \psi>\mid \ge E_g,$$ 
we get
$$
E\ge \hbar \Omega+\frac{N(N-1)}{2}E_g +g^{\ast}\mu_b B S_z
\eqno{(15)}
$$
where $E_g$ is the ground state energy of the effective two-body Hamiltonian 
$H_{12}$ (for the completely spin polarized states, $S_z=N\hbar/2$ and $E_g$ is the energy
 of the lowest antisymmetric state of the effective two-body Hamiltonian $H_{12}
$).
Eq.(15) is a generalization of the Hall-Post method (which is restricted to the case with only interparticle forces present and no external potential) for obtaining lower bounds 
 to the ground-state energy of N-electron QD in a magnetic field $B$.
 
\vspace{18pt}
{\bf III. Large $B$ Limit}

\vspace{8pt}
We introduce dimensionless units by making the following transformation: $
\vec{\rho}=(1/a) \vec{\zeta}_1$, where $a=\sqrt{\hbar/(m^{\ast}\omega_0)}$.

Using the above dimensionless notation and polar coordinates
$\rho_x=\rho \sin\theta$ and$~$ $\rho_y=\rho\cos\theta$, we can write the effective 
two-body eigenvalue problem,\\ $H_{12}\mid \phi_g>=E_g\mid \phi_g>$ as
$$
\tilde{H}_{12}u(\rho)=[-\frac{2}{N}\frac{d^2}{d \rho^2}+\frac{2(\ell^2-1/4)}{N \rho^2}+
\frac{1}{2N}(1+\frac{\lambda^2}{4})\rho^2+2\frac{\gamma_c}{\rho}-\frac{\ell 
\lambda}{N}]u(\rho)=\tilde{E} u(\rho),
\eqno{(16)} 
$$
where
$\lambda=\omega_c/\omega_0$, $~$$\phi_g(\rho,\theta)=e^{i\ell \theta}u(\rho)/\sqrt{\rho}$, $~$$\gamma_c=\alpha \sqrt{m^{\ast}c^2/(\hbar \omega_0)}/2$, and
$\tilde{E}=E_g/(\hbar \omega_0)$. 

The two-body equation, Eq.(16), can be solved numerically  
 to find $E_g$ for any
arbitrary value of $\ell$. The optimal $\ell$ value, restricted to odd integers
for polarized states, minimizes the energy. The two-electron QD
has been the subject of intensive study [3,4,8,22-26]. 

In the large magnetic field limit, $\ell$ becomes large and 
 the term $1/4$ in $(\ell^2-1/4)
$ can be neglected [17], and  Eq.(16) can be rewritten as
$$
[-\frac{2}{N}\frac{d^2}{d \rho^2}+V_{eff}(\rho)]u(\rho)=\tilde{E}u(\rho),
\eqno{(17)}
$$
where
$$
V_{eff}(\rho)=\frac{2 \ell^2}{N\rho^2}+\frac{1}{2N}(1+\frac{\lambda^2}{4})\rho^2+2\frac{\gamma_c}{\rho}-\ell N\lambda.
\eqno{(18)}
$$
In the large-$B$ limit the effective potential $V_{eff}$, Eq.(18), has a deep 
minimum, therefore a good approximation to  $\tilde{E}$ can be obtained by
making 
the Taylor expansion of  $V_{eff}$ about its minimum [17]. Thus the approximate 
$E_g$ is
$$
E_g\approx \frac{2}{N}[\frac{3}{4}(2\gamma_c N)^{2/3}+
\frac{1}{2}\sqrt{\lambda^2+3}]\hbar \omega_0.
\eqno{(19)}
$$

Substitution Eq.(19) into Eq.(15) gives
$$
E\ge E_{low}\approx {\cal{E}}=\hbar \Omega+(N-1)[\frac{3}{4}(2\gamma_c N)^{2/3}+
\frac{1}{2}\sqrt{\lambda^2+3}]\hbar \omega_0+g^{\ast}\mu_BBS_z.
\eqno{(20)}
$$
Note that the large $N$ limit of $E_{low}$ is 
independent of magnetic
 field  in this approximation (see also Ref.[16]).

\vspace{18pt}
{\bf IV. Numerical Results }

\vspace{8pt}
We begin with the single-electron basis functions $\chi_{n\ell}(\rho)$, associated with Hamiltonian
$$
H_0=-\frac{d^2}{d \rho^2}+\frac{\ell^2-1/4}{\rho^2}+\frac{1}{4} 
(1+\frac{\lambda^2}{4})\rho^2-\frac{\lambda \ell}{2}.
\eqno{(21)}
$$
These functions were found more than seventy years ago [27],
$$
\chi_{n\ell}(\rho)=A_{n\ell}\rho^{\ell+1/2}e^{-(1/4)(1+\lambda^2/4)^{1/2}\rho^2}
L_{n}^{\ell}((1/2)(1+\lambda^2/4)^{1/2}\rho^2),
\eqno{(22)}
$$
where $L_n^{\ell}$ are associated Laguerre polynomials  and
$$A_{n\ell}=[\frac{1}{2}(1+\frac{\lambda^2}{4})^{\ell+1}\frac{n!}{(n+\ell)!}]^{1/2}.
\eqno{(23)}
$$
In order to solve Eq.(16) we introduce
$$
\chi_{n\ell}^{\beta}(\rho)=\sqrt{\beta}\chi_{n\ell}(\beta \rho)
\eqno{(24)}
$$
and expand $u(\rho)$, Eq.(16), in the basis $\chi_{n\ell}^{\beta}$,
i.e. we seek solution of the form
$$
u^{M}(\rho)=\sum_n^M c_N^{\beta}\chi_{n\ell}^{\beta}(\rho).
\eqno{(25)}
$$
The conventional choice for the parameter $\beta$ is $\beta=1$ (see, for example [4]). However, for finite $M$, the choice $\beta=1$ is not the optimal choice.
The most reliable $\beta$ is obtained from
$$
\frac{d}{d\beta}<u^M\mid\tilde{H}\mid u^M>=0.
\eqno{(26)}
$$
We apply the method, Eqs.(25) and (26), to compute the lower bounds.

Consider a two-dimensional three-electron QD with $\epsilon=13.1$ and $\hbar\omega_0=0.01$ meV without a magnetic field, B=0 [18]. Let M be the number of functions in Eq. (25). Examples of the lower bound to the completely spin polarized three-electron state, $S_z=3\hbar/2$, corresponding to the different $M$ are given in Table I. The fast convergence is evident. Comparision of the converging result of Table I, $E_g=0.336659$ meV with exact diagonalization calculations of Ref.[18], $E_g=0.3393$ meV, shows that our lower bound is a very good approximation with relative error of about 0.7\% for the ground state energy of the three-electron QD without a magnetic field.

Now consider the $GaAs$ QD with $\epsilon=12.4$ and $\hbar \omega_0=4$ meV in a
strong
magnetic field, $B=20$T [19]. Examples of the lower bounds to the completely spin polarized N-electron ground state, $E_{low}$, for up to $N=6$ electrons are given in Table II. Numerical results $E_{low}$  agree with large $B$ approximation, $\tilde{E}$ (Eq.(20)) to better than $0.1\%$.

From Table II, we can see that the calculated lower bounds agree well with 
exact-diagonalization results, $E_{ed}$ [19]. The relative error,
$\Delta=(E_{ed}-E_{low})/(2 E_{ed})$, is less than $2\%$.

 We have also calculated the chemical potential of QD, $\mu_A(N)=E(N+1)-E(N)$.
$\mu_A(N)$ is measured by the single-electron capacitance spectroscopy method [20, 28] 
since the transfer of the $(N+1)$th electron from the electrode to the QD occurs when 
the chemical potential of the electrode, $\mu_E$, is equal to the $\mu_A$. 

\vspace{18pt}
{\bf V. Summary and Conclusion}

\vspace{8pt}

 In summary, we have generalized the Hall-Post  lower-bound method [21] for
the case of the $N$-electron parabolic QD in the presence of a perpendicular magnetic field.

It is shown that our rigorous lower bounds agree well with the results of exact diagonalization. For example, lower bounds to the completely spin polarized ground state energy of the three-electron QD agree with exact-diagonalization results to better than 1\%. For the case of six-electron QD, the relative error
is less than 2$\%$.

Analytic results for the lower bounds to the ground-state energies of the QD in a strong magnetic field (the QD analogue of a Wigner crystal [29] known as electron molecule [19]) agree with numerical lower bounds  to better than 0.1$\%$.

\pagebreak

Table I. Convergence of the method, Eqs.(25-26) for lower bounds, $E_g$ with increasing $M$ for the three-electron QD with $\epsilon=13.1$ and $\hbar \omega_0=
0.01$ meV. $S_z=3\hbar /2$ and $B=0$ is assumed [18].\\ 

\begin{tabular}{lllllll}
\hline\hline
$M$
&1
&4
&5
&6
&7
&8\\ \hline
$E_g$, meV
&0.373368
&0.336831
&0.336681
&0.336659
&0.336659
&0.336659 \\ \hline\hline
\end{tabular}\\

\vspace{18pt}

Table II. Results for lower bounds $E_{low}$, chemical potential 
$\mu_A$, large $B$ analytical approximation ${\cal{E}}$, Eq.(20),
and $\Delta=(E_{ed}[19]-E_{low})/(2 E_{ed}[19])$ for N-electron $GaAs$ QD with $\epsilon=12.4$ and $\hbar \omega_0=4$ meV in a strong magnetic field, $B=20$T.
$S_z=N\hbar /2$ is assumed.\\

\begin{tabular}{lllll}
\hline\hline
Number of electrons, $N$
&$E_{low}$, meV
&$\mu_A$, meV
&${\cal{E}}$, meV
&$\Delta$, $\%$\\ \hline
1
&17.4810
&24.2141
&17.4810
& \\ \hline
2
&41.6951
&28.5087
&41.6910
& \\ \hline
3
&70.2038
&32.0032
&70.1483
&0.2 \\ \hline
4
&102.207
&35.1570
&102.168
&0.7 \\ \hline
5
&137.364
&38.2630
&137.366
&1.4 \\ \hline
6
&175.627
&40.6922
& 175.482
&1.8 \\ \hline \hline
\end{tabular}	 

\pagebreak

\begin{center}

{\bf References}
\end{center}

\vspace{8pt}

\noindent
[1] M. A. Kastner, Phys. Today {\bf 46}, 24  (1993).

\noindent
[2] M. A. Read and W. P. Kirk, {\it Nanostructure Physics and Fabrication}
(Academic Press, Boston, 1989). 

\noindent
[3] L. Jacak, P. Hawrylak and A. Wojs, {\it Quantum Dots} (Springer, Berlin, 1998).

\noindent
[4] T. Chakraborty, {\it Quantum Dots} (Elsevier, 1999).

\noindent
[5]  D. Bimberg, M. Grundmann and N.N. Ledentsov, {\it  Quantum Dot 
Heterostructures} (John Wiley \& Sons, 1999).

\noindent
[6] N.A. Bruce and P.A. Maksym, Phys. Rev. B{\bf 61}, 4718 (2000).  

\noindent
[7] G.W. Bryant, Phys. Rev. Lett. {\bf 59}, 1140 (1987).

\noindent
[8] D. Pfannkuche, V. Gudmundsson and P.A. Maksym, Phys. Rev. B{\bf 47},
2244 (1993).

\noindent
[9] H.R. Post, Proc. Phys. Soc. A{\bf 66}, 649 (1953).

\noindent
[10] N.F. Johnson and M.C., Payne, Phys. Rev. Lett. {\bf 67}, 1157 (1991);
N.F. Johnson and M.C., Payne, Superlattices and Microstructures {\bf11}, 309 (1992). 

\noindent
[11] B.L. Johnson and G. Kirczenow, Phys. Rev. B{\bf 47}, 10563 (1993).

\noindent
[12] S. Gartenhaus and C. Schwartz, Phys. Rev. {\bf108}, 482 (1957).

\noindent
[13] R.B. Laughlin, Phys. Rev. B{\bf 27}, 3383 (1983). 

\noindent
[14] P.A. Maksym and T Chakraborty, Phys. Rev. Lett. {\bf65}, 108 (1990).

\noindent
[15] P. Hawrylak and D. Pfannkuche, Phys. Rev. Lett. {\bf 70}, 485 (1993);
P. Hawrilak, Phys. Rev. Lett. {\bf 71}, 3347 (1993).

\noindent
[16] S.-R. E. Yang, A.H. MacDonald and M.D. Johnson, Phys. Rev. Lett. {\bf 71}, 3194 (1993).

\noindent
[17] P.A. Maksym, Phys. Rev. B{\bf 53}, 10871 (1996).  

\noindent
[18] X.-G. Li, W.-Y. Ruan, C.-G. Bao and Y.-Y. Liu, Few Body Systems {\bf 22},
 91 (1997).

\noindent
[19] P.A. Maksym, H.Imamura, G.P. Mallon and H. Aoki, J. Phys.: Condense Matter {\bf 12}, R299 (2000). 

\noindent
[20] R.C. Ashoori, H.L. Stormer, J.S. Weiner, L.N. Pfeiffer, K.W. Baldwin and 
K.W. West, Phys. Rev. Lett. {\bf71}, 613 (1993).

\noindent
[21] R.L. Hall and H.R. Post, Proc. Phys. Soc. {\bf90}, 381 (1967).

\noindent
[22] M. Wagner, U. Merkt and A.V. Chaplik, Phys. Rev. B{\bf 45}, 1951 (1992).

\noindent
[23] M. Taut, J. Phys. A:Math. Gen. {\bf 27}, 1045 (1994).

\noindent
[24] P.A. Maksym, {\it From Quantum Mechanics to Technology} (Springer Notes in Physics, {\bf 477}) Eds. Z. Petru, J. Przystawa and K. Rapcewicz (Springer,
1997) p. 23.

\noindent
[25] M. Dineykhan and R.G. Nazmitdinov, Phys. Rev. {\bf 55}, 13707 (1997).

\noindent
[26] B.A. McKinney and D.K. Watson, Phys. Rev. B{\bf61}, 4958 (2000).

\noindent
[27] V. Fock, Zeitschrift f\"ur Physik {\bf 47}, 446 (1927);
C.G. Darwin, Proc. Cambridge Philos. Soc. {\bf 27}, 86 (1930).

\noindent
[28]  R.C. Ashoori, H.L. Stormer, J.S. Weiner, L.N. Pfeiffer, K.W. Baldwin and
K.W. West, Phys. Rev. Lett. {\bf 68}, 3088 (1992).

\noindent
[29] E.P. Wigner, Phys. Rev. {\bf46}, 1002 (1934).
\end{document}